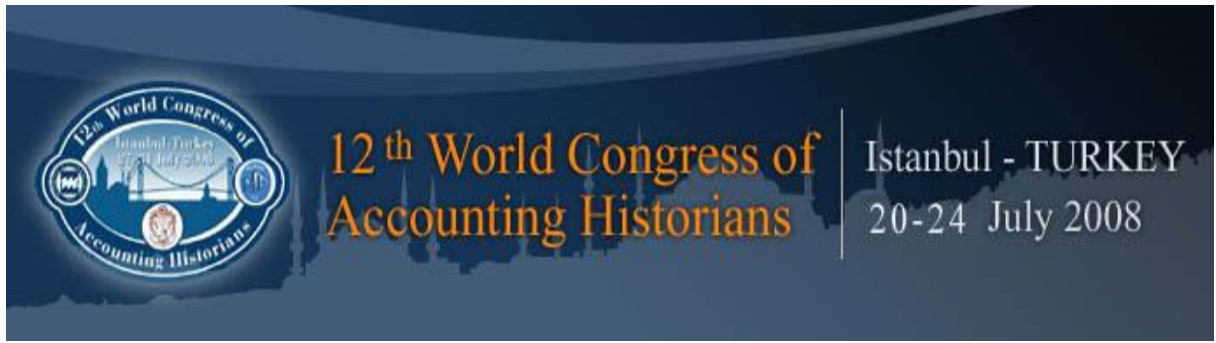

# A discussion of stock market speculation by Pierre-Joseph Proudhon

___________________________

*Nice February 2008*
_________

**Jean-Claude JUHEL**
**Professeur des Universités**
**Université de Nice - Sophia Antipolis**
Groupe de Recherche en Management de l'IAE de Nice
GRÉMAN (ex CRIFFP) EA 1195 / UNSA
*Ancien Directeur de l'Institut Universitaire*
*de Technologie Nice - Côte d'Azur*

**Dominique DUFOUR**
**Maître de conférences des Universités**
**Université de Nice - Sophia Antipolis**
Groupe de Recherche en Management de l'IAE de Nice
GRÉMAN (ex CRIFFP) EA 1195 / UNSA
*Professeur à l'Institut d'Administration*
*des Entreprises de Nice*

# A discussion of stock market speculation by Pierre-Joseph Proudhon

*Abstract*

The object of this contribution is to present the ideas behind the thinking of the French economist Pierre-Joseph Proudhon (1809-1865) in relation to the causes and effects of Stock market speculation. It is based upon the works of this author but particularly on his "*Manuel du spéculateur à la Bourse*" (Stock Market Speculator Manual) edited in 1857 in Paris.

Compared to the markets of today, however, the stock market described by Proudhon appears embryonic. Nevertheless it represents the location for transactions in financial assets, commodities, precious metals and even some transactions involving options.

This contribution is organised in the following manner - the first section is devoted to the development of Proudhon's thought in relation to speculation. It is divided into two parts. The first part is dedicated to Pierre-Joseph Proudhon's definitions of stock market speculation or gambling with shares that for him served no purpose either from a human or economic perspective and was therefore condemnable and to be contrasted with entrepreneurial speculation that, even though it is a highly-risky activity, involves the spirit of enterprise and provides the lifeblood of economic growth. The second part allows us to present Pierre-Joseph Proudhon's propositions in relation to restricting the speculation that he considers obnoxious.

The second section has two objectives: one part places in perspective the views of Proudhon and the characteristics of stock market activity under the Second Empire whilst the other part examines current-day aspects of the characteristics evoked by Proudhon. We are interested especially in the question of the regulation and that of the relevance today of certain accounting practices.

________

# A discussion of stock market speculation by Pierre-Joseph Proudhon

On the 15th of December 1856 Pierre-Joseph Proudhon explained within the preface to the third edition of his *"Manuel du spéculateur à la Bourse"*[1] (Stock Market Speculator Manual) his decision to abandon the anonymity under which he published the first two editions. Upon receipt of an order from the publisher Garnier Frères, he originally thought that the publication of a *"compilation of stock market transactions"*[2] did not merit his signature. One could not but be surprised that an ardent proponent of a socialist republic during the Second Empire would write a users' guide intended for stock market operators; one can, however, understand the discretion with which his political orientation constrained him from producing a publication that could be accused of popularising the subject matter for the mass market.

Beginning with the third edition, Proudhon intends to consider doctrines on the stock market or the *"temple of speculation"* that derive from his political and social writing. He wants to assess stock market speculation in relation to general social standards or *"public morals"* and to *"economic development"* of his era. The unrestrained speculative development during the first half of the 19th century is for Proudhon simply a reflection of a corrupted social and economic system and a forewarning of an insurmountable barrier for the economic mechanism that should deliver a just and sustainable society - *"the industrial republic."*

The object of this contribution is to present the ideas behind the thinking of the French economist Pierre-Joseph Proudhon (1809-1865) in relation to the causes and effects of Stock market speculation. It is based upon the works of this author but particularly on his *"Manuel du spéculateur à la Bourse"* edited in 1857 in Paris.

Compared to the markets of today, however, the stock market described by Proudhon appears embryonic. Nevertheless it represents the location for transactions in financial assets, commodities, precious metals and even some transactions involving options.

This contribution is organised in the following manner - the first section is devoted to the development of Proudhon's thought in relation to speculation. It is divided into two parts. The first part is dedicated to Pierre-Joseph Proudhon's definitions of stock market speculation or gambling with shares that for him served no purpose either from a human or economic perspective and was therefore condemnable and to be contrasted with entrepreneurial speculation that, even though it is a highly-risky activity, involves the spirit of enterprise and provides the lifeblood of economic growth. The second part allows us to present Pierre-Joseph Proudhon's propositions in relation to restricting the speculation that he considers obnoxious.

The second section has two objectives: one part places in perspective the views of Proudhon and the characteristics of stock market activity under the Second Empire whilst the other part examines current-day aspects of the characteristics evoked by Proudhon. We are interested especially in the question of the regulation and that of the relevance today of certain accounting practices.

---

[1] P.-J. PROUDHON - *Manuel du spéculateur à la Bourse*, Garnier frères, Paris, 1852 ; this first edition was published anonymously. Idem, 3ème edition, 1857, considerably enlarged and signed (in collaboration with Duchêne, former editor of the newspaper *Le Peuple*). Idem 4ème édition, 1857.

[2] The expressions or citations in italics and in brackets are taken from the work of Proudhon, the subject of this study.

1. The nature of speculative activity

Proudhon's stated objective for his, "summary of stock market data," is to "*serve the interests of anyone who may be compromised whilst not attempting to hide the vagaries of stock market fluctuations*," - pensioners, shareholders, landowners, merchants, heads of families as well as "*all those whose capital is invested there*" via public funds, industrial enterprises and via property. The value of their assets changes according to the daily movements of the exchange; "*for everyone…we win or lose each day something at the Stock exchange…*" in terms of capital, interest or in relation to the price of commodities or of raw materials. He sees society under the Second Empire, dominated by the Stock market as a vast, economic battlefield "*…where the means of production act as a combat weapon…*", "*…where everybody is forced to adhere only to the rules of war…*" "*…morals are characterised by dishonesty, business is dominated by piracy.*" This assertion applies to all the financial centres of the period: London, Paris, Vienna, Hamburg, Frankfurt and Amsterdam. The principal cause of this is stock market speculation, which Proudhon defines directly as "*the collection of means, ignored by the law and thus inaccessible to creditors seeking compensation, that permit the assets of others to be taken*" and thus representing something far more significant than a simple game of chance.

We will look at Proudhon's treatment of stock market speculation in his manual first and then at his description of stock market activity under the Second Empire.

A discussion of stock market speculation by Pierre-Joseph Proudhon

Proudhon represents speculation as a type of production. However, like all human activity it can be misused – "*unproductive speculation,*" the stock market acting as, "*the temple,*" for this ritual. He adds that unproductive speculation is an inherent part of the economic system.

He provides several examples of speculative transactions in order to show the reader the economic necessity of one of them - "*productive speculation*" – a high-risk activity but one which is the motor of economic growth, and to demonstrate the economic perversity of the other – "*the unproductive*" – a risky activity but also a sterile one in economic and human terms and so is guilty of encouraging all those factors that leads the economic system to its ruin. Gambling in general and stock market speculation in particular constitute the basis of unproductive speculation

1.1.1 Speculation is one of the four forms of production

His definition of production remains heavily influenced by the ideas of the classical school. He oscillates between the idea that only human economic activity is capable of nurturing economic growth (Ricardo) and that which attributes such productive properties also to capital (Jean-Baptiste Say).

He therefore states that the production of economic wealth is the product of human activity and it thus derives from labour although this process can take one of four forms: labour, capital, trade and speculation.

But however it is expressed, Proudhon uses the "classical" idea that economic production is not concerned with the creation of raw material but the creation of utility: all activity that adds value to raw materials, by transformation, delivery, transportation, etc is productive.

1° Labour is the transformation of raw material by the hand of man. Manual labour of all sorts defines the "*working class.*" However labour only becomes productive via synthesis with the intermediate factors of land and of capital. Land is the first form of capital supplied free to man by nature.

2° Capital derives from accumulated labour. The product of labour becomes capital when it is acquired by another producer. "*Capital and labour merge into one another.*" Thus, worked land becomes capital. Over time, the accumulation of capital concentrates itself in the hands of the capitalist or owners' class. Lending (leasing, rent, tenant farming, granting exploitation rights, investment companies, etc) is the process by which capital passes from the hands of the capitalist or owners' class to those of the labourer or industrialist. The investment company with shares is the extreme example of this transfer; the public company does not really change this: the shareholders actually being the providers of funds for the use of the elected Directors.

Proudhon takes the idea involving the productive combination of factors of production: capital and labour (Turgot, Smith). No factor is productive alone. They do not become so until they are combined with talent – by an entrepreneur—or with genius by a productive speculator.

3° Trade is the transport or movement of merchandise without which such goods remain valueless. Within the activity of the exchanges, Proudhon includes banking transactions with an emphasis upon the circulation of money. Any activity that adds value is for him productive.

4° Finally, speculation is the fourth economic actor that is involved in production. It represents, for Proudhon, the intellectual conception of the different procedures by which labour, capital and trade can feature in production. Once again, any activity is productive that satisfies a need.

Speculation looks for the "*buried riches*" via new processes of transformation, of lending, of transport, of trade or of the circulation of capital. Its nature is thus random as any intellectual conception that expects to be justified by experience. "*The productive speculation*" is the source of progress and provides orientation for labour, capital and trade.

He cites as an example the financing of high-risk activities (during his era): maritime trade, fire insurance, marketing, technological research. Today, these activities would be categorised as innovative or as innovations whilst the term "speculation" itself appears immoral.

He also describes the founding principle of Law's system whereby metallic monetary transactions are substituted by those involving only paper and expresses his regret that this innovation – productive speculation – had been attacked by "*crazed gambling*" – or unproductive speculation.

When he concludes that "*speculation is... the genius of discovery…that invents, innovates…that creates something from nothing,*" this relates undoubtedly to innovation, a term that could easily replace Proudhon's "*productive speculation.*" Innovation "*always on alert, with inexhaustible resources, sceptical of prosperity, ever present in the dreams…announce, conceive, rationalise, define, organise, order, finalise; labour, capital and trade execute.*"

This description excludes "*unproductive speculation.*" However, the abuse of speculation should not be confused with its errors. The error of speculation results from misjudged anticipation or from unfortunate miscalculations (this was the case for Law's system or other innovations without prospects).

By contrast, three examples of speculation in relation to raw materials are not convincing in relation to their productive influence since in each case all that is involved is speculation on a rise in the price of commodities (wood, wine and wheat) and not on a source of progress: Even if in each of these situations there is no wish to deceive and if, *in extremis*, there is a service provided to the consumer – thus the speculative hoarding of wine allows the market to be supplied when all other supplies are exhausted – there is effectively a speculative hope that prices will rise. Some current commentators admit, however, like Proudhon, that in such circumstances speculation plays a positive role: it encourages an optimal allocation of available resources.

Unproductive speculation can be criticised in a moral context and destroys the system via its corruption of principles. Throughout his work he insists upon the necessary connection between justice and economic activity: *"Justice and commerce must…be interdependent...justice serving as commercial law.*[3]*"* Within an economic context unproductive speculation diverts capital from productive use.

1.1.2 Speculation leads to abuse: the unproductive speculation

All industrial, commercial and financial combinations are governed by chance: speculation thus involves some level of risk. The reward from the transaction is therefore two-fold: for the newly created utility there is the reward for a service and a speculative gain for the risk assumed. This second part, the jobbery, is the object of abuse: sought for its own sake, irrespective of any service supplied, this remuneration of risk can be classified as a game or even as a fraud.

Speculation is thus the art of becoming rich *"without working, capital, trade or genius."* Proudhon did not imagine risk being rewarded for its own sake, unrelated to a speculative activity: this would be *"illicit and immoral"* since it involved the direct appropriation of the assets of public funds or individuals. It is in addition in conflict with any return on capital, detracting revenue from the working classes without providing anything productive.

Most stock market speculation is based upon either random calculations or upon restricted information.

The first group of speculations includes games and betting in all their forms, particularly the purchase and sale within the account that is often performed in precarious circumstances: purchase without the available funds when the trade is opened and sale without possessing at that date the contract for the asset concerned.

The second category relates to *"misdemeanours or commercial crimes: misrepresentation, deceit, monopoly, cornering a market, misappropriation of funds, disloyalty, extortion, fraud, theft."* Proudhon was very shocked to find that *"the financial world admits, tolerates, excuses or pardons such acts;"* he provides the reader with no less than fourteen examples of speculation, taken from real life, illustrating these different forms of theft being supported by illicit intent, the abuse of influence, diplomatic intrigue, lies, intimidation, legal transgressions, hoaxes and favouritism.

If these errors are solidly anchored in the selfishness of human nature, it is the economic system that encourages their expression and their development. The abuses that accompany production and above all speculation are the principal reason for the formation of the antagonistic classes: the bourgeois and the proletariat.

---

[3] De la justice dans la Révolution et dans l'Eglise, Edition Rivière, 1930, tome III, p. 60

### 1.1.3 *"Unproductive speculation"* is an inherent part of the economic system

Commercial existence shows that there exists extensive "*dependence and solidarity*" to varying degrees between all speculative business. It must be remembered that speculation dominates the entire economic system since it underlies and drives work, lending and trade. Every rise or fall of asset values consequently affects the entire economic system (stock market, banks, companies, society) since all economic transactions are interconnected.

For Proudhon, the economic social system attains its peak of perfection when labour, capital, trade and speculation are shared fairly between all citizens; on the contrary, it is sub-optimal when these activities are appropriated by particular classes that then act as a distinct group. Hence, some classes exploit others. The political system is thus the reflection of economic organisation, which explains the latter's evolution and disappearance.

In particular, more and more individuals take advantage of and profit from unproductive speculation to the detriment of all of society: "*morals are characterised by dishonesty, business is dominated by piracy.*" Proudhon argues therefore that the current regime at the Stock exchange and of the banking system is causing mortal harm to the economic system. He demonstrates that financial transactions without a productive basis "*lead fatally, in the current state of affairs, to fraud and theft.*" Two categories of causes explain this casualty: the "*complicity of science and of the law*" on the one hand, and the "*unequal role played by the participants*" on the other.

a) Complicity of science and the law:

In an economic system that is based upon individual behaviour and not upon conformity, that is to say, it is based upon, "*the absence of mutual interest amongst the factors of production*" no law prohibiting abusive speculation is possible.

The period of time between delivery and payment, being the condition of lending and of trade, produces a risk to production and to the transportation of goods. Proudhon observes that, as a consequence of the prominence of random chance in the mechanics of the economy and due to the lake of the sharing of risks, gambling becomes the norm: "*Collective interests act to limit randomness (insurance) whilst gambling acts to accentuate it.*"

In these circumstances incompatible with the principle of justice, no economic theory can legitimise the futures markets.

It is evident that Proudhon is more of a doctrinaire than a theorist: science is not concerned with morals and only studies facts. Society is a product of circumstances; its transformation is a political and not a scientific problem.

He adds that gambling diverts money from productive applications and produces, via the workings of the stock market, a meaningless assessment of the use of capital: the official accounts of the service of market-specialists in Paris recorded 80 million francs in commission during the year 1850; according to Proudhon, this implies a total of transactions of 64 billion of which, he evaluates that only 6% relate to serious business – so much capital is thereby diverted from a productive destination.

For most of the unproductive transactions, the majority are based upon the inequality of the situation of the participants.

b) The inequality of the situation of the participants:

Irregularities in the transactions "*add explicitly to the immorality of the game*" Proudhon emphasises. He completes the picture of the participants on the stock exchange: the gamblers, market manipulators, the swindlers in exchange transactions and in transactions involving limited partnerships.

1 – The gamblers.

The stock market public is divided into *"exploiters"* and the *"exploited."* This second category is by far the most numerous. The exploited play the stock market in the same way as they play the lottery. The exploiters who live off the proceeds of the prior category, can be divided into the *"prudent"* and the *"skilful."*

The *"prudent"* perform arbitrage transactions throughout the year. Arbitrageurs profit from the momentary differences of prices or rates, without taking risks and wait for these imbalances to disappear. Proudhon criticises them nonetheless: *"They consider themselves to be eminently useful citizens and voluntarily express their indignation at and criticise gambling and earning a living as a parasite...the newspapers....the manuals...the almanacs of the stock exchange portray them as being exemplary."*

Today arbitrage is not considered as a speculative activity to the extent that it promotes equilibrium in the markets. Even though he understood this aspect, Proudhon comments, rather thoughtlessly no doubt, that, *"if there were only serious transactions"*, there would not be any imbalances to harvest. From an ethical perspective, arbitrage is for Proudhon an *"unproductive speculation"*.

These are the *"clever, combined with the needy that set the game in motion and the prudent who maintain it."*

The *"clever ones"* with little or no capital, speculate everyday. They play without funding their positions, profiting from differences without ever owning the subject of the transaction. Their activity is not without risk but bankruptcy remains the exception.

2 – The market manipulators.

The market manipulator influences the market to his advantage. In order to modify the game involving the exchange transactions of independent participants, an *"association of capital and intelligence"* is required.

Above the large or small gamblers one finds the combination of wealth and intelligence: using their joint power they *"orchestrate"* the market.

Each speculator who wins believes themselves to be clever even though he is only playing a game of chance, albeit in the context of high finance.

Understandably, high finance has its dependants (doorman, mistresses, friends, etc) that it will forewarn of favourable opportunities.

3 – The fraudsters.

Proudhon describes two types of fraudulent activity:

- Cheating involved in exchange transactions
- The broker speculates against his clients
- Employees of the state profit from their position
- The captains of industry bet against the shareholders, against the creditors and against their company.
- Collusion between the specialist newspapers and the speculators.

- Cheating involved in the transactions of limited partnerships.
These are the various and fraudulent financial transactions carried out by the founders and/or directors of investment companies.

Proudhon concludes his analysis of speculation with a moral indictment against the stock exchange, the activity of which corrupts public morals. The stock exchange is the pulse of society and reflects the morals of the country. The influence of the stock market has spread everywhere. Its spirit dominates the whole of Europe. Finance exploits every aspect of society.

The eradication of speculative activity

The second part allows us to present Pierre-Joseph Proudhon's propositions in relation to the eradication of this speculation that he considers so obnoxious. According to Proudhon speculation is inherent to the economic system, spontaneous, unconstrained and limited only by the imagination of the participant: because of this, it is impossible to regulate.

In this regard, Proudhon asks himself in what ways a prohibition of the essence of speculation, forward trading, would be possible. He states that only a controlled and directed economy would allow this, in eliminating the oscillations of supply and demand. The state would control production, prices, output and consumption. However, *"to remove all the risky aspects of production, of the transportation and consumption of goods…would be to remove the factors that stimulate the entrepreneurial spirit."* Proudhon concludes that such a controlled organisation is not compatible with the pursuit of economic and social progress.

In the circumstances prevailing at the end of the first half of the XIX century, according to Proudhon society began an *"unrelenting march towards a state of crisis."* A social revolution that would establish *"mutuellisme"* (mutualism) would be able to prevent the devastating social effects produced at the peak of the crisis. Proudhon remained realistic regarding the voluntary and deliberate establishment of collectivism, however: public opinion does not embrace it and the government will not take the initiative. The negative economic and social consequences of the abuse of speculation, he hopes, must lead – despite everything – to *"industrial democracy."*

The inexorable march towards a crisis

Social injustice, class antagonism and social conflict, the appropriation and concentration of capital, speculation and the deterioration in moral standards are the characteristics of the three phases of the approaching crisis: *Industrial anarchy, feudal industrial production and the industrial empire.* Proudhon clearly adopts the role of moralist in the analysis of this development.

He intends to show that France of 1850 attained the ultimate degree of feudal industrial production. The pursuit of economic concentration will lead the country to a situation that culminates in a crisis – that is to say *"to the systematic capitalist and speculative absorption"* to the industrial empire. This stage, in maximising social conflicts, will trigger the revolution that leads to *"industrial democracy."*

1- *Feudal industrial production: progress of the crisis.*

Feudal industrial production has today replaced the industrial anarchy that was bequeathed by the Revolution of 1789. It is characterised by the systematic adoption of the granting of privileges and by the acceptance of immoral transactions that profit a group of perpetrators who thereby extract close to half of total national wealth.

The impoverishment of the *"working"* class that characterises the current crisis of the system is attributable, according to Proudhon, to several causes:

- Random causes: calamities of all kinds that limit national production but that could be anticipated via a *"mutualistic"* insurance system.
- Inherent causes related to the way our society is organised:

    * The exorbitant importance of capital invested in machine tools, particularly in railways, that is drawn to the stock market; Proudhon emphasises the extent to which this squandered; from 1830 to 1855 the net revenue withdrawn did not increase in the same proportion as the capital committed.

    * Over time, the workforce became remunerated less, to the detriment of the self-employed and of small industry.

    * The rise in the cost of living, following the diversion of agricultural products to the towns, facilitated by the railways, etc that was to the detriment of the rural population and, consequently:

    * The redirection of agricultural finance towards the stock exchange leading to the disappearance of smallholdings, permitting the formation of large farms and favouring the conversion of arable land to pasture,

    * This, in turn, produces a depopulation of the countryside, and:

    * An impoverishment of the soil poorly exploited, which explains the periodic instance of bad harvests.

    * A rise in the level of rents, due to the rural exodus, to the benefit of the class of small property owner parasites.

    * Continual growth in taxes due to an uncontrolled increase in expenses of the state that represents, in 1855, 20% of national product.

    * An ever increasing need for money in circulation, normally required for economic activity, to facilitate stock market speculation - i.e. sterile transactions on securities - which paralyses labour and trade.

    * A decline in the moral standing of the country: abandonment of productive work in favour of unproductive speculation.

2- *The industrial empire: peak of the crisis*

In the absence of a deliberate wish to pursue reforms, developments themselves will force economic and social renewal. The excesses of the industrial concentration that constitutes "*the industrial empire*" will lead to a destruction of the system.

An upper class controls society whilst representing only 1 in 80 numerically and subjects the other two classes, middle and lower, to abuse: their interests converge. The middle class fell progressively into a precarious condition. The annoyance of the burden of taxation, the distributions of capital and of property always concentrated in fewer hands, the development of large enterprises that crush small undertakings and relegate the middle classes from *"the exercise of professional occupations to a subordinated employee positions."* The working class, the most numerous, finds itself at the level of slavery, poverty and even greater levels of insecurity. Their situation is thus connected: *"The circumstances of well-being that the former require assume the achievement of those claimed by the latter and visa-versa. The worker would have work guaranteed if the bourgeois himself has a guarantee of entrepreneurial success; the consumer would find a low cost of living if the producer is able to rid himself of the parasites that stop and hinder him…"*

To attain "*mutuellisme*"[4]

Proudhon thus argues for *"a radical transformation of society towards liberty, equality between people and towards a confederation of people..."* But he does not support violence or destruction. However, can one expect to remove privileges, unearned income or the concentration of power…without strongly imposing such a change? Proudhon refuses to accept any responsibility for a possible catastrophe, considering his role as only that of a kind of *"prophet"*: labour *"having found the secret of financing itself"* no longer accepts an oppressive economic regime. The worker must be joint-owner of capital and of the resulting profit and not a slave.

To achieve this peaceful revolution it would suffice, according to Proudhon, to redistribute property, realised in an equitable manner via *"a simple process of repayment"* that he calls *"liquidation"* – a term that is nevertheless ambiguous and that today has, through the history of radical developments, acquired a clear sense of human brutality.

When capitalism is liquidated, in the sense of Proudhon - that is, in an accounting sense – *"industrial democracy"* will follow.

1 – The liquidation of capitalism: general repayment.

Proudhon only sees one issue arising during the crisis: *"liquidation"* in an accounting sense – the repayment of capital.

The annual profits will be calculated in two parts: one in relation to dividends and the other in relation to the repayment of capital. At the end of a certain period, the length of which is unimportant Proudhon says, the capitalists' investments are repaid and expropriated.

After the *"liquidation"* of capital has begun, repayment becomes an indispensable process for all loans. The redemption of debts implies another structure for credit, notably an unspecified reduction in interest in order to render the repayment simple. Interest disappears or, at least it is limited to the payment of the management costs involved in making capital available.

The *liquidation* of existing companies involves:
- For the railways, canals, mines, insurances, banks, etc - so for the collection of the industrial and financial sector – the replacement of the investment vehicles of the capitalist class by collectively-owned industries and associations of workers...
- For the business of trade and of exchange, to abolish the monopoly of the institutions and all the privileges of intermediaries; to discourage the inclination to gamble via powerful guarantees from specifically created national and regional institutions and, in this way, to create a vast system of advertising, balance and control in order to remove any opportunities for unproductive speculation.

2 – Industrial democracy – financing of labour by labour or universal mutualism: end of the crisis.

Proudhon advocates an *"economic reconstitution…mutualism."* There is a mutualistic spirit within a company when all the workers are supposed to work for each other. Similarly, these workers associations will be united according to the same principle of mutuality. This self-management structure is based on the principles of liberty, transparency and above all the reciprocal support of each other. According to Proudhon it must eliminate all the injurious behaviour (theft, fraud, extortion, etc) and all of the economic imbalances (over production, stagnation, unemployment, etc).

---
[4] Author's note: word coined by Proudhon which meaning is similar to "mutualism".

The new society addresses four preoccupations, the solutions to which render speculation impossible. The production of wealth will be assured by the *"workers associations,"* the satisfaction of the population's needs by *"consumers associations"* and borrowing requirements by *"swap associations."* In relation to the accommodation for workers, this will be organised in a way that avoids completely the mistakes of the *"workers' housing estates."*

\* Workers associations

Workers associations will replace the current private sector companies where workers and shareholders are exploited. They should not be mistaken for the communist production units that do not respect the rights of the individual. They will be characterised as follows:
- Participation of all the associates in the management and profits of the enterprise.
- An appropriate structure combining piece work and salaried work
- A pension and sickness insurance scheme funded via deductions from salaries and profits
- Progressive education of apprentices
- Collective guarantee of collaboration: this means supplying intermediary and consumer goods at the lowest possible price between the associations.
- Publication of the accounts.

Proudhon cites seventeen associations that would have operated at this time in Paris according to these principles (jewellery, workshops, masonry, coach-building, etc). He notes, however, that a number of other workers associations established at the time of the Second Republic (bakers, road workers, hatters, etc) have disappeared, *"overcome by adversity…by the strife…all have born the cost of their inexperience…"*

One will notice that the real legal nature of the production units used as an example by Proudhon is not specified; are they not simply the self-employed enterprises or small to medium industries, created by self-employed associates with modest initial capital and employers of workers that they sensitively designate as *"auxiliaries"* who perhaps participate in the distributed profits? It does not seem even to be the case that one could equate them with what one will later call producer co-operatives.

Besides, except for the system of co-operative agriculture, in relation to which the entrepreneurial spirit is not really perceptible and of some consumers' co-operatives or again of mutual insurance, where the interest of the consumer never seems to be a priority, history does not provide such self-managing initiatives whatever their scale as shown by the Yugoslav system.

\* Consumers associations

This does not relate to the Consumers associations that we have today. Although rather unclear in his proposal, Proudhon appears to describe and hope for the creation of commercial units for retail distribution, that acts like a small supermarket but which are organised like a consumers' cooperative: *"– a coalition between consumers, who guarantee a commercial organisation their clientele and a profitable existence…and…in return, receive a discount to current prices in respect of the produce from this organisation…"*

He declares that some examples of *"consumers associations"* (bakers, butchers, herbalists) had begun to establish themselves in the main locations of specific areas, thanks to *"the financing of several –enlightened– bourgeois"* but following a coup d'état on the 2$^{nd}$ of December these projects miscarried although Proudhon does not explain why.

* The swap associations

The swap associations must revolutionise the economy since Proudhon envisages the withdrawal of money that is the instrument of unproductive speculation. Money, dressed-up by the interests of capital, represents *"a fortuitous right"* of the owner to collect a part of the product of the workers: interest, discount bills, commissions, tenant farming, rent, added-value, etc

Everybody must be able to exchange immediately the product of his work for the product of somebody else's work.

In this way the *"parasitic organisations"* that obstruct the distribution and development of trade would disappear. For this, Proudhon proposes the formation of swap associations facilitating the general usage of vouchers that are valuable for all consumable products and are issued by the producers in place of money who will then consent to receive them in exchange for their products. Established without capital by a number of managers of the organisation, for which the products serve as collateral, its trade banks, having no capital to remunerate, can lend without charging interest. It is sufficient to receive a simple commission covering the cost involved in the transaction.

As capital is made freely available to all in the form of vouchers, since money disappears, there only remain the workers who exchange their products at cost prices. *"Exploitation disappears, justice is assured, class differences are eradicated and Government …becomes superfluous.[5]"*

As one could imagine, the futures markets disappear in these circumstances. It must be acknowledged, however, that the structure of such a project is not very convincing. There is no sense in undertaking a detailed critique of the project, however, since, as has already been stated, Proudhon sees all this rather from the perspective of a moralist than that of an economist or manager. Thus, the abolition of interest does not depend upon any banking mechanism but rather upon the good faith of the participants. As everybody is aware, interest represents the remuneration for a temporary relinquishment of capital to a third party, together with a risk premium.

If the criticism of the poor living conditions of the lower class, of the decline of the middle class and of the abuse of speculation, the significance of which derives from the expropriation of capital involved, contains an irrefutable aspect of truth, the reconstruction of society as proposed by Proudhon is unrealistic in many ways. It seems that Proudhon has a utopian vision of humanity in general and of his contemporaries in particular. Man seems naturally good and honest if the economic system does not prompt him to behave in a corrupt manner. He does not seem to realise that it is the behaviour of men that shape the economic system. Business regulation in general and accounting regulation in particular – in relation to the futures markets, for example - will only act, according to him, so as to discourage initiative: he proposes individual liberty within a mutualistic and transparent regime.

He says nothing, however, that is very precise about the effective management of the associations that he describes as examples; nothing either in relation to the management rules that would have to govern such mutualistic enterprises. The reader understands that it involves self-management but remains ignorant of how this would work on a day-to-day basis. Proudhon recognises that it is not the sharing of the profits that is the driving force for this, because these become insignificant at the level of the worker. Individual motivation resides, according to him, in the mutualistic system that includes insurance against life's adversities although these notions are also explained without much detail.

---

[5] Organisation du crédit et de la circulation… Guillaumin, Paris, 1848 (Ed. Lacroix t. VI, p. 121)

He brings no more precision concerning the mutualistic relationships between the workers associations. Proudhon seems to imagine that an "invisible hand" will push people towards the initiative, towards cooperation, coordination, economic progress, solidarity and entrepreneurial brotherhood.

Proudhon without illusion

*"But this fortuitous revolution appears unlikely to come to fruition…"* as public opinion does not adopt it, even in relation to those who have been wronged; in so far as enlightened observers are concerned – like Proudhon – they appear as utopians or revolutionaries. The governments adopt no initiative for reform. Consequently, it is necessary to wait until the excessive speculation itself stimulates a mutation of the economic system.

The French bourgeoisie missed an historic opportunity with the Revolution of 1789 that should have inspired them. The institution of stock exchanges imposed upon them a three-fold task:
- In relation to themselves: promote free enterprise and meritocracy whilst preventing *"speculation that is pure gambling…"*
- In relation to the workers: promote the growth of well-being (fair salaries, local authority housing and profit-sharing) and of education…
- In relation to the Government: promote the redemption of debt and contest favouritism in all of its forms.

*"For an intelligent bourgeoisie, generous and upright, the stock exchange could have been Parliament, from which decrees would come each day that are more efficient than all the Orders ...and laws…"* Unfortunately, since the birth of the monarchy of July, parasitism and gambling dominate. Without differentiating productive from unproductive speculation, the working class classifies all bourgeoisies as their class enemy. This antagonism appears decisive to Proudhon.

The economic evolution of the first half of the XIX century is, despite everything for Proudhon, a period of hope. He sees in the speculative abuse, the hope of renewal leading to a renaissance of humanism, to an equitable society, founded upon economic mutualism and the political aspects of federalism[6] of justice and social peace. Federalism also guarantees peace between different peoples.

Proudhon completes the workings of the industrial democracy by describing the ultimate mechanism of the production of wealth and of economic development. Of course, unproductive will be transformed into productive speculation. The capitalist released from sterile and alienating speculation, will engage his capital in innovative businesses that will remunerate and pay him back at the same time. *"Invent now, make discoveries, construct machines, and create, with new needs, new products..."* Labour, *"– inexhaustible source of finance –"* will act as a guarantee here.

Proudhon addresses his comments to speculators – the readers of his manual – and attempts to convince them that this new system will be favourable to them in liberating the working class from its alienation: *"form companies, in anonymous and collective names; obtain for your successful combinations, for your useful applications, for your sustainable*

---
[6] We will not be treating in this study the political ideas that compliment "collectivism": "federalism."

*enterprises patents and franchises, distribute your shares around; undertake business of millions and billions and do not worry…"* you will be remunerated and repaid by work; *"the redemption of debts combined with the productive power...of the companies of workers...will offer an unlimited facility to create wealth..."*

Dispense therefore with economic anarchy, speculative individualism, the exploitation of man by man that benefits *"a single and the same association of producers."*

Despite the movements for change in society such as alterglobalists, equally diverse as the interests that they serve, a radical social change of this type today still seems utopian. The imposition of rules is considered as the means of regulating the financial markets and of controlling speculation.

2. The limits and current relevance of Proudhon's work

The stance of Proudhon in relation to the functioning of the stock exchange is ambiguous. His work testifies to his admiration of the ingeniousness of its participants and it recognises its social role in a liberal regime. At the same time, he prays for its disappearance. History has not supported his stance. This argument allows us to consider the stock exchange of Paris in the XIX century, so as to specify what the characteristics are that could lead one to an equally critical description; and this when, as an instrument of financing, the stock exchange will have played a crucial role in France's development during this century. We will also examine two aspects of stock market speculation during the XIX century: regulation and accounting practices. In relation to this second aspect we consider current practice in respect to the issues raised by Proudhon.

The stock exchange in Paris during the Second Empire

The development of the Stock exchange of Paris and, to a lesser degree, the provincial exchanges, was noticeable from 1850 onwards. It was driven mainly by four types of activities: the construction of canals, of living accommodation, of railways and finally due to banking. The point must be made that at this date there already existed a futures market as well as a market for the exchange of derivative instruments, where options where traded. These observations derive from the voluminous study: "*Le Marché français au XIX° Siècle*" (Hautcoeur, Gallais-Hanonno and alii, 2007).

Participants and markets

There follows a description of the participants and markets at the time of Proudhon. Two points must be raised here relating to two of the principal characteristics of this market: the Coulisse (Curb market) and the futures market are illegal. As Gallais-Hannono wrote, "*the stock exchange of the XIX century has two characteristics: the more dynamic part of the exchange was an absolutely illegal black market: The Curb market and the majority of transactions made on the official exchange were also themselves illegal – these are the forward transactions.*" (Gallais-Hannono, 1996)

1. The participants

The operation of the exchange requires the participation of market intermediaries. These market intermediaries are of two types: The officials and the officious.

The officials unite the market-specialists and the brokers. A market-specialist is the holder of a franchise. They alone have the power to trade the quoted securities and to establish their prices. He is responsible for the payment for and for the delivery of the securities traded. A broker is an intermediary that operates between a client and a market-specialist. He can take the orders for the client and act for his account but the trading of securities is always performed by the delegated market-specialist. Neither a market-specialist nor a broker can in any case act for their own account within the context of stock exchange transactions. Commercial and banking activities are prohibited for them. These professions are organised and there exists a syndicate chamber of market-specialists charged with the organisation of the profession in respecting the legal constraints.

The officious intermediaries are called unlicensed brokers and curb agents. These intermediaries act wholly illegally but this illegality is tolerated by the public powers. There thus exists in fact a market as well as a system of quotation that one could classify as official, created by unlicensed intermediaries. What advantages does the Stock exchange gain from the presence of the Curb market? They are numerous: the Curb market ignores the trading hours of the exchange, thus ensuring a continuous quotation; the transaction costs are lower and finally, they provide specialist introductions for securities ignored by the market-specialists who concentrate their attention upon the placement of government securities. These advantages are accompanied by a higher level of risk to investors by comparison to those involved in the official market. It should be added that the Curb market has allowed the establishment of a type of liberal market: a market consecrated to the sale of securities of unquoted companies. The companies active on this market could hope to accede one day to an official quotation. However, it must be added that in relation to the primary market, the issuance of securities is unconstrained and competitive and so the services of a market-specialist are not required in these circumstances. The Curb market goes on to become gradually more official and it will end up merging with the Compagnie des Agents de Change (market-specialists' company) in 1962.

2. The markets

When Proudhon was writing, there existed on the stock market of Paris the following markets: the cash market and the futures market.

1. The cash market

This market is that involving payment in cash and the delivery of securities within a maximum period of 5 days following the trade date.

2. The futures market.

The status of the futures markets is, at the time when Proudhon was writing, ambiguously described. The Civil Code prohibits any legal action relating to a bet or to a gaming debt. Forward transactions, appearing to fall into one of these categories, have for a long time been prohibited by the justice system on the basis that, *"the futures markets having no objective except of differences, must be classified as exchange games and cancelled as futile and unrealistic, as contrary to the law, to public order and morals."* (Proudhon, op cit page 78). But in 1856 the Imperial Court of Paris admitted their legality subject to the qualification that *"the vendor justifies, by regular offers, having had in his possession on the expiry date, the number of shares sold by him."* (Proudhon, op cit page 79) Forward transactions are thus tolerated albeit with an uncertain legal status. The maximum deferral period accepted is of one month for railway shares and two months for other securities. It should be noted that there exists the possibility of prolonging these periods by carrying-over a position. Short sales are possible here. In other words, the seller is not obliged to be in possession of the securities when a carry-over is agreed. Within the market exist two compartments: the firm market and the premium market.

On the firm market, the undertakings to buy and to sell at an agreed price are fixed. The settlement day occurs monthly and the market-specialist can demand a margin payment. This margin payment is less than the value of the securities in question. There thus exists a leverage influence on this firm market.

The premium market corresponds to what one would call today an options market. As in the contemporary system, the buyer has the right and not the obligation to buy. In 1856 there did not exist any options to sell but only options to buy. The counterparty to this right is the payment of a premium.

Also as today, the impact of premium transactions via exercise is uncertain. Proudhon comments on page 86: *"the futures markets are the real battlefield of speculative gambling."* At the same time he contradicts himself – this is common throughout his work - when he admits that the premium market is less risky than the cash market. He writes thus in talking of the main futures market operating without margin. *"The enormity of the risks of this type of operation prompts one to imagine another market that is less injurious and in respect of which we are going to expose the workings (this concerns the premium market)."* (Proudhon, op cit page 84).

Proudhon always oscillates between a moral condemnation on the one hand and recognition of the benefits deriving from certain of the mechanisms that he describes on the other. He thus writes, *"Unfortunately, to restrict the futures markets would be nothing less than to restrict trade"* (Proudhon Op cit page 87). On this point, Proudhon is clairvoyant since he senses well that the options market allows the futures market to function – its presence is necessary taking account of the uncertainties of business life - at a lower cost than that of the main section of the futures market.

In addition it is possible to combine the operations on the firm market with those on the premium market. These transactions are undertaken for speculative purposes, their combination allowing a modification of the risk involved to a level that suits the investor.

As for today's markets, the majority of transactions effected on the premium market do not lead to an exchange of the underlying securities. Proudhon estimates a relationship of 15 to 1 for the volumes traded on the premium market and those on the firm market, both cash and forward transactions. This ratio is impossible to verify as the reconstruction of reliable statistics is today extremely difficult.

The settlement days here are also monthly. The possibility to carry-over a position is available to those who would not be in a position on this date to exercise their undertakings. A law of 1885 will render the futures market entirely legal.

Corporate and capital structures

We limit ourselves here to private sector securities, leaving to one side Government securities. The corporate structures most frequently used are the "société anonyme" (Public Limited Company) and above all the "société en commandite par actions" (Investment Trust). The latter is more often – more so than the former - subject to an imperial license. At a time when the limitation of the responsibility of associates is coming to the fore, the investment company appears to be a form that protects individual shareholders more. It should be added that the legislator has become aware since this time of the necessity of protecting shareholders against possible abuses of the directors. Even though the structure of the investment company requires a strong sense of responsibility on the part of the latter, the law monitored the implementation of protective mechanisms that we could call governance procedures. Thus, from 1850 the presence of a supervisory council charged with an examination of the accounts, of a verification of the inventory, of an analysis of the dividend and of the organisation of meetings was rendered necessary.

The opportunity created by the creation and development of these companies gave rise to three types of securities:
1. Shares reserved for the founders of the company. They are created when the company is formed. Their price is usually lower than those at which the other shares are issued. They are often called industrial shares. They do not pay interest to holders.
2. Ordinary shares They combine the characteristics of both equities, as they contribute to the capital of the company and of bonds as they offer a fixed coupon, often guaranteed by the Government and are repaid via a random redemption process - one should be aware that for the railways, the company, *a priori*, has a limited life equivalent to the duration of the concession. Capital must be repaid progressively.
3. Participating shares carry a right to dividends when a specified period has elapsed. This period can correspond to the repayment period for all of the shares that act as a loan or to the completion of certain work.

The coexistence of these three types of securities lies at the core of the practices that Proudhon condemns. Effectively, it carries the seeds of conflicts of interest that the final section (below) will allow us to illustrate.

The subscription to a share when a company is introduced to the exchange is not accompanied necessarily by the sale of all of the capital. As a general rule only 25% of the nominal amount of the shares must be made available on the date of the initial offering. There is then more potential for a bullish performance on the stock market.

Sectors of activity that dominate the exchange

We now consider the three types of corporate activity that dominate the quotations during the XIX century.

1. The railways

The development of the exchange in the XIX century is mainly due to the emergence of the railway companies. Only the stock exchange could satisfy their needs for financing. The system usually adopted in order to define the context of the activity was that of the concession. In this system, the partial financing and then the management are entrusted to the companies subject to the meeting of a certain number of obligations. The undertakings of the Government and of the companies vary from one railway company to another. We should note, however, that the concession does not lead to an indefinite franchise being granted to the concessionaire. This type of operation appears much more risky for the savers than that of the submission of a tender bearing in mind that the raising of initial funds is the work of private companies and not of the Government. It should be added that the concessions are not for an indefinite term. Their duration is limited. 30, 45 or 99 years. In these circumstances, the companies are under an obligation to schedule the repayment of the capital over the period of the concession.

2. The canals

The investments in the railways and, to a lesser degree, in the canals constituted the most important infrastructure operations of the XIX century. In the system of tenders, utilised for the canals, the Government takes responsibility for planning, the realisation as well as the operation of the canals. The financing is via a private counterparty and thus the company retains a right to a proportion of the user tolls. This system reflected the coexistence of shares relating to a loan and those relating to equity in the company. The low level of income from the canals forced the Government to buy the equity shares, under pressure from private tender companies, in order to avoid their closure in several cases (Nieradzik in Gallais-Hanonno and Hautcoeur, 2007).

3. The banks

We leave aside the Banque de France, of which the status was already rather unique during the XIX century. There exist three different types of banking establishments.
- The Discount houses. These establishments discount bills and loans on the basis of the collateral of merchandise to which they relate. Since 1854 their financing is provided via public savings.
- Mortgage lenders. The most important of which is Crédit Foncier de France, created in 1852 and funded by public savings. The statutes foresee a prudent management approach, particularly with: an absence of dividends in case of losses, the establishment of a reserve fund, the size of which should reach 50% of initial shareholders' funds, followed by further subsequent financing from the issue of bonds or equities.
- The securities and investment organisations. Société Générale de Crédit Mobilier is the prime example of such an organisation. This is a general bank that also allocates time to make selected participations in other companies. Its financing is also principally via the issue of debt. The degree of leverage can be significant: the debt issues can amount to 10 times the size of shareholders' funds. As Proudhon notes: *"If ever the bourse falls, the public collateralised debt and company shares* (that constitute part of the depreciating assets) *and the share capital will begin to be reduced."* (Proudhon, op cit p 258). The conditions of the usage of the profits are similar to those of Crédit Foncier de France.

The significance of the stock exchange: commercial and political

The stock market in Paris was officially recognised in 1801. The monopoly over transactions thereon is conferred upon the market-specialists. The development of the securities exchanges during the XIX century in France is spectacular. Three periods can be identified - 1800-1850: slow progression, 1850-1880: rapid progression and finally 1880-1900: instability. In one century the relationship between stock market capitalisation and GDP grew to close to 50% (Arbulu, 1998). This phenomenon was sustained until the war of 1914-1918. Actually, in 1913 this ratio is close to 75%. At this date, Great Britain is the only country for which this ratio exceeds that for France – for the USA, this ratio was shrinking. This level is remarkable and following the falls that accompanied the two world wars, it was not until 1995 that this ratio returned to this level in France (Rajan and Zingales, 2001). The stock exchange will have therefore occupied a role more significant during France's economic growth of the XIX than during the XX century.

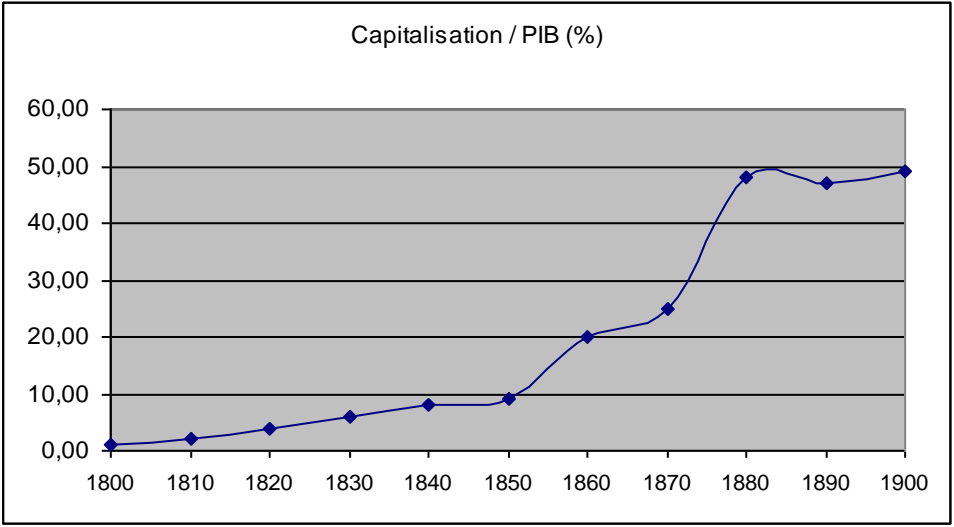

The relationship between market capitalisation and GDP in France during the XIX century
(Source: Arbulu. P, 1998)

It is remarkable to witness, beginning in 1850, the emergence of stock market placements as a tool of wealth management in competition with traditional financial investments. This growth has several causes: economic growth, of course but also the weakness of the Government's resources meant that it had to rely upon the private sector to finance infrastructure construction that was necessary for this growth. By infrastructure is meant: works, equipment, public services and facilities (Arbulu and Vaslin, 1999). The financing of this prime infrastructure, amongst which features the railways and canals, in the absence of sufficient bank finance, depended upon the stock market. The Government's role takes the form of agreements with the private companies so as to profit from their competition and to satisfy the investment needs (Arbulu et Vaslin, 1999).

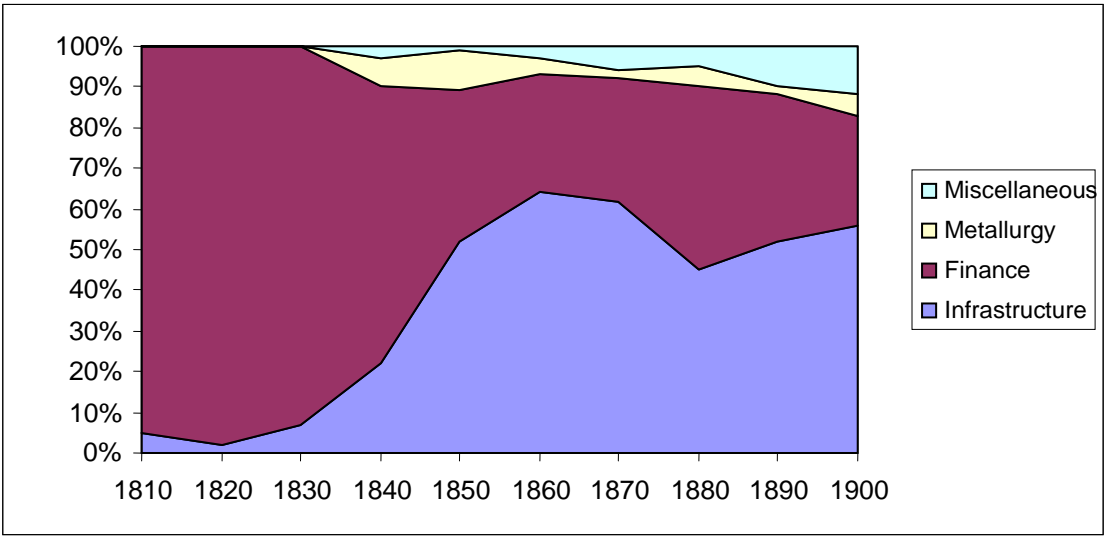

Stock market capitalisation by sector
(Source: Arbulu and Vaslin, 1999)

The French stock market landscape in the second half of the XIX century was thus characterised by: savings being strongly attracted to the financial markets, capital intensive industries developed – especially the railways – and finally, banks heavily engaged in financing appeared, both short- (discounting) and long-term via long-dated loans. We should add that some of these banks act as intermediaries and thus serve to transfer risk.

The second half of the XIX century is going to be recognised for a long bull trend composed of cycles involving correctional declines. The question of returns from stock market investments over the XIX century is of course difficult to categorise. G Gallais-Hanonnno estimated an approximate annual return of 5% from shares in canals and railways of the XIX century (Gallais-Hanonnno, 1996). Government securities returned something similar. Due to the provision of guarantees for interest and then to the support given to companies in difficulties, the Government reduces the risk to which equity investors are exposed to thereby allowing them to facilitate a similar return to that available from the government's own securities.

### Speculation, governance and accountability

Proudhon has identified a number of practical circumstances that fuel speculation. We will restrict ourselves here to the examination of two of these that remain relevant today: the weakness of the corporate government of quoted companies on the one hand and questionable accounting practices on the other.

### Speculation and governance

The development of the Parisian stock exchange during the XIX century is accompanied by the emergence of questions relating to corporate governance in those companies where ownership is not concentrated in a single person. Proudhon caught a glimpse of the problems that this access to capital could produce. For France, the issue is not that of a possible conflict of interest between managers with little ownership in the company and the shareholders, it rests rather in the opposition between major shareholders - who are often the company founders - and minority shareholders. Proudhon illustrates two of these: the representation of the minority shareholders and the price of the shares.

#### 2.2.1.1 An insufficient representation of the shareholders

Corporate governance practices during the majority of the XIX century showed themselves to provide insufficient protection for the shareholders. This lack of protection appears characteristic, in any case, for the French model by comparison with the Anglo-Saxon governance models. This relates to the participation of the shareholders at the general meetings. In the XIX century restrictive clauses were often implemented. Thus, the ownership of a minimal quantity of shares could be demanded.

For Crédit Foncier de France, the principal mortgage lender, participation at the general meetings was restricted to the largest 200 shareholders.

Other measures restrict the right of shareholders to put forward points for discussion at the general meetings. So, in the case of Société Générale de Crédit Mobilier, only those propositions signed by at least 10 shareholders who have the right to participate at the general meeting could be discussed.

French capitalism at this time was characterised by a strong concentration of the control of companies within the hands of a limited number of actors.

2.2.1.2 Capital transactions preceding the introduction to the Exchange.

Before the introduction to the Exchange, the founders of the company derive capital from their own investments. Proudhon points out that these investments are made at a value below that at which the company is introduced. There thus exists a bonus that benefits certain shareholders. Hence, in the case of the Versailles railways, for the initial offering shareholders subscribed at 700 francs per share whilst for a subsequent augmentation of the capital the price was 725 francs.

These bonuses denounced by Proudhon continue to attract discussion and diverse opinions. The COB (French Stock Exchange Regulatory Body), issued a communiqué in 2000 on this subject entitled "Communiqué: Transactions in the capital of companies in the weeks before an introduction to the exchange: the COB seeks consultation from the professionals" (COB, 2000). In the introduction to this communiqué the COB wrote: "*For several months, the Commission has noticed that a significant and growing proportion of the initial offerings were preceded, within a period of less than one year beforehand, by one or several increases of capital at prices considerably below the initial offering price. These operations often involve investment funds, possibly those linked to the introducing bank. They often accompany the granting of options or of vouchers allowing employees of the company being introduced to access the capital at a price that has no bearing with that subsequently offered to investors. This development poses a two-fold problem: the equality of treatment between different classes of shareholders as well as the transparency of initial offerings, particularly where such dilutions to the share capital occur during the period when the documents are being presented to investors who do not always have the possibility of including the incidence of such changes to their calculations*" (COB, 2000, p 11).

The retrospective examination of certain transactions is interesting in this regard. Thus, for an example here the society Auféminin.com should be cited. This company was founded on the $14^{th}$ of July 1999. The nominal value of its shares is €20. On the $17^{th}$ of December 1999, the capital was increased via cash subscriptions at a single price of €24.48. On the $31^{st}$ of March 2000 the nominal share value was divided by 100; the price paid for a share then became €0.2 and €0.2448 for the first and second wave of shareholders respectively. The company was introduced on the exchange on the 19th of July 2000 at a price of €7.60. The initial offering price is therefore 38 times the amount paid by the founding shareholders less than a year beforehand. The relationships illustrated by Proudhon seem uninspiring by comparison with such a return. Before implementing a guideline destined to address these practices, the COB orchestrates a consultation process amongst those involved in initial offerings (COB, 2001). Subsequently, new rules were announced by the COB: no capital transaction will any longer be possible from the moment when the documentation is given to its services and information "*complete and analytic*" in relation to all capital changes that have occurred during the three years prior to the admission to a quotation must be included in the offering documentation. Of course, some exceptions will arise but capital transactions that occur during the period during which subscriptions to the initial offering can be made can only be done so at a price that is no less than 20% lower than the price for the initial offering.

Speculation and accountability

It is interesting to note that some of the accounting practices denounced by Proudhon continue to pose problems today. We will examine three of these here: The regime for depreciation, the capitalisation of costs and the payment of dividends deducted from shareholders' funds. Actually, they all arise from the non-respect of the principle of fixed nature of capital. The question raised is that of the existence of a real distributable profit and in a broader context, to include the definition of the notion of capital itself, by which here is meant shareholders' funds. We begin by a question more relevant at this time: the maintenance of a "capital account." We will restrict ourselves to the railways, the emblematic sector in relation to the Exchange of the XIX century. Since the beginning of the XIX century the law recognised two principles that happened to be linked: The prohibition of fictional dividends and the principle of the fixed nature of capital (Lemarchand et Praquin, 2005). However, the absence of an established, uniform system of accounting renders their implementation difficult.

2.2.2.1 The maintenance of a "capital account."

One of the first practices denounced by Proudhon is the "*maintenance of a capital account.*" Here we cite the text that describes the financing of the railways "*an important point was the definitive closure of the capital account. In many companies, this account remained permanently open, even though the railway activity that the company undertook had finished a long time ago. This system of accounting allowed the directors to continually allocate to the capital account all the costs of improvements that scientific and transport engineering development suggested were necessary. Some expenditure that should have remained in the revenue account, were diverted and added to the capital account, in order to allow the distribution of dividends that maintained the confidence of the shareholders and mislead the public in relation to the value of the companies*" (Proudhon, p 161).

The two questions alluded to in the quotation are of a slightly different nature. The opening of the capital account is to enable the implementation of infrastructure work that is in addition to that initially agreed in the articles of association of the company. In these circumstances, in forcing the company to resort to a loan in order to finance these works, the repayment of a part of their own funds and the payment of dividends becomes more difficult. It is necessary to know that at the time of which Proudhon talks, funds raised had to be used in relation to the specific nature of the investment for which they were intended.

The introduction to the Exchange and the increase of the capital that accompanied it had as an objective the raising of funds necessary to launch the company. These funds thus had to be of a sufficient amount in order to deal with all of the needs of the company. This was not the case, however, if the expenditure required turned out to be more than the amount of equity funds raised. Thus Foucaud demonstrated in his thesis a gradual progression, beginning in 1840, of maintenance expenditure and renewal expenditure on the part of the British railways (Foucaud, 2006).

The practice consisting of "*leaving the capital account open*" is also indirectly linked to the phenomenon of the financial bubbles following the crisis of a downturn. The introductions on the Exchange during the XIX century were characterised by the raising of funds from shareholders but also from the banking system in the form of loans. As such, they were different to current initial offerings as these are usually for much more significant amounts than the short-term needs of the company require. Bank loans thus no longer seem necessary for the completion of the financing. In the case of the railways it is remarkable that the development of the companies is accompanied by a constant growth in the proportion of debt in relation to total financial resources. Capital thus continued to be fed via lending following the Introduction. A. Joanne cites the following figures: in 1850 the equity funds raised represented 80% of the cost of the project. This proportion was only 43% in 1854 (Joanne p 11). The accounts of the Chemins de Fer du Nord supplied in the annex provide an illustration.

The second technique – that one generally refers to as the capitalisation of costs generally accompanied by under-depreciation – has the effect of enhancing the result of the company.

2.2.2.2 Under-depreciation

The first questionable practice is that which consists of not depreciating assets. Depreciation here is intended to mean the allocation of costs to the revenue account reflecting the deterioration in the value of fixed assets. In not depreciating assets, the company enhances its results and thereby facilitates the distribution of dividends. This effectively leads to a consumption of its assets and thus of its own capital. The case study of the Chemins de Fer du Nord referred to in the appendix also supplies an example of this practice.

2.2.2.3 The capitalisation of costs

This involves a technique that has often been used so as to massage results. Some companies involved in the Internet bubble resorted extensively to this type of practice. It has even occurred that the cost of providing meals has been capitalised on the basis that it represented the creation of an asset. It is interesting to note that the opening of the capital account and the capitalisation of costs can be accompanied by a lack of depreciation. One could imagine that the former compensates for the latter. Depreciation can be conceived as the recognition of the falling value of assets but also as a means of maintaining the productive assets. In this second perspective, the opening of the capital account and the capitalisation of costs would allow for the under depreciation to be compensated for. This view is mistaken as although the assets are affected, it is actually the liabilities that are more involved here. Capitalise costs and under-depreciate acts to disguise the consummation of shareholders' funds. This remark serves as a link to the third questionable accounting practice.

2.2.2.4 The non-respect for the rule that capital should remain fixed.

The third practice removes what one has called the principle of fixed capital. It involves a legal principle originally that gradually has become an accounting principle. This principle arises with the limitation of responsibility of the associates. As N. Praquin writes: "*The knowledge of the capital pledged by the associates is an important aspect involved in the protection of creditors, given that responsibility is limited. To the extent that this capital is limited to the founders' investments, it becomes of fundamental importance to insure that its value is preserved in the company. This is the reason that regulatory guidance requires adherence to the principle of fixed capital; as it is better both not to discourage the shareholders and also, thereby, to be in a position to assure them of dividends, even in the absence of profits*." (Praquin, 2005, p 276). The context is thus special since it also relates to guidance for the investment companies; at the same time, however, it is necessary to monitor the protection of shareholders' interests in commercial companies. The implementation of this principle is not easy to the extent that it is precisely articulated with that of the independence of the annual accounts and within the context of a business continuity assumption. In relation to the question that interests us there is a practical problem: the distribution of statutory interest to the shareholders. This issue is summarised by Proudhon as follows: "*Companies that require significant funding and that are involved in works lasting several years, such as the railways, pay interest to investors on a regular basis. However, when the operation has* *not produced any profits, these interest payments can only be taken from capital.*" (Proudhon, p 205).

As this legal principle is well established, how is it that its contravention by companies can arise? The opaqueness of the accounting practices combined with a weakness of the regulatory organisations would have contributed to lapses (Praquin, op cit). At the same time, the Government, having suggested that the public should save, was able to turn a blind eye in order to encourage savers to invest in companies.

The combined effect of these lapses did nothing to undermine the success of the Stock exchange of Paris in the XIX century. One simple reason explains this: the sustained role of the Government. It would be tempting to compare the development of the financing of the railways with the internet bubble of the end of the 1990s. Of course, similarities exist but the fundamental difference resides in the participation of the government. In guaranteeing the payment of statutory interest and in intervening in the case of difficulties for any of the quoted companies, the Government has assumed the business risk on its own account. This encouraged all the participants to accept all the *a priori* questionable practices, particularly those involving accounting techniques. It should be noted that the two themes that we have discussed: the question of corporate governance and that relating to the quality of accounting information remain more relevant than ever today. The continual efforts of different organisations to make progress on these subjects serve to illustrate that all that is produced are the conclusions of *ad-hoc* committees in relation to good corporate governance or to the comprehensive implementation of new accounting standards or new organisational structures. The objective should be to put an end to speculative manoeuvres that could be based upon reprehensible corporate governance and accounting information.

Conclusion

With the benefit of hindsight, the work of Proudhon appears to us to be ambiguous and limited. First of all it must be noted that there is a clear contradiction between the adoption of a position opposed to the Stock Exchange and the publication of a speculator's manual. Then we would remark that this opposition in principle hides a certain admiration for the energy of the speculators as well as the recognition of their social utility, notably in forward transactions that are the most heavily criticised in other contexts by the author. Proudhon undoubtedly blackened the influence of speculation by emphasising certain practices that were condemnable in his view and this was almost certainly because at the time the French Exchange accommodated practices that were legally prohibited. They were even necessary to its operation – we are speaking here of the Curb market and of the futures market. Proudhon was undoubtedly disturbed by the significant role played in the development of the Exchange of certain capitalists ever present in the banks and in the railways. Proudhon also highlighted questionable accounting practices. Since the time when he was writing, an enormous effort to standardise accounting practices has been and continues to be made. The absence of a uniform system in the XIX century did not, however, restrict the growth of the Exchange. The presence of the Government as guarantor of a minimum shareholder return via the frequent granting of a guarantee for the interest payments represented a powerful encouragement to invest on the exchange and led the shareholders to close their eyes to the often oblique accounting practices. This governmental presence encouraged, at the time of Proudhon's writing, a strong correlation between the political environment and the dynamism of the stock market. Developments in the world of politics could cause large movements in stock market valuations.

Today the government no longer assumes this role. The importance of the quality of accounting information is thus crucial; the establishment of standards IAS is, in any case, characterised by concerns reaffirmed many times and aimed at improving information destined for the external investor.

**Appendix**
**Chemins de fer du Nord 1857**
**Accounting information dating from 1857 (the numbers are in thousands of francs).**

| Financial situation | | | |
|---|---|---|---|
| **Assets** | | **Liabilities** | |
| Construction expenditure incurred by the Government | 80 785 | Funds paid on old shares | 160 000 |
| Expenditure on materials incurred by the Government | 3 073 | Funds paid on new shares | 25 000 |
| Remaining to pay on the construction | 5 230 | Loan of 1852 | 24 750 |
| Interest owing to the Government | 10 810 | Loan of 1854 | 22 989 |
| Constructions | 302 110 | Loan of 1854 | 22 429 |
| Treasury | 31 349 | Loan of 1855 | 21 556 |
| Payments remaining on the new shares | 46 875 | Loan of 1856 | 21 850 |
| | | Loan of 1857 | 20 817 |
| | | Bond issue | 37 500 |
| | | Bond issue | 1 181 |
| | | Debts in relation to the Government | 16 035 |
| | | Trade creditors | 5 585 |
| | | Interest and dividends owing | 28 639 |
| | | Track maintenance funds | 6 439 |
| | | Reserve and depreciation | 18 587 |
| | | Payments remaining on the new shares | 46 875 |
| **Total** | **480 232** | **Total** | **480 232** |

| Cash flows | | | |
|---|---|---|---|
| **Application of funds** | | **Source of funds** | |
| Constructions | 37 128 | Treasury at the beginning of the period | 24 300 |
| Payments to the Government | 2 000 | Payments received from the new shares | 25 000 |
| Treasury at the end of the period | 31 349 | Payments received from the loan of 1857 | 20 817 |
| | | Diverse | 360 |
| **Total** | **70477** | **Total** | **70 477** |

| **Expenditure** | | **Receipts** | |
|---|---|---|---|
| Administrative charges | 1 080 | Passenger transport | 20 623 |
| Operating charges | 7 257 | Freight transport | 34 047 |
| Cost of materials | 7 186 | Discounts | -4 380 |
| Work on the track | 3 459 | | |
| **Total** | **18 982** | **Total** | **50 290** |
| | | | |
| **Profit** | **31 308** | | |

| **Distribution of profits** | |
|---|---|
| Interest of the shares less revenue from investments | 6,025 |
| Depreciation of the capital | 233 |
| Interest and repayments of loans | 6,546 |
| Interest on funds owed to the Government | 187 |
| Track maintenance funds | 360 |
| Pension fund | 165 |
| Reserves | 179 |
| Dividends | 17,600 |
| Carried forward: | 13 |
| **Total** | **31,308** |

Here is the accounting information of a railway company relating to 1857. The document comes from the work of Joanne (Joanne, p 24, 1859).

Several characteristics should be commented upon. In so far as the commercial aspects are concerned, two elements are remarkable: the gross return: 31,308 from a turnover figure of 50,290 and the consolidated amount of financial costs due to lenders (6,546) and to shareholders (6,025 + 17,600 + 233). So far as the nature of the accounting information presented, the detail of the presentation should be noted (consisting of a review, a profit and loss account, a cash flow analysis and the distribution of profits); the terminology is modern since the IASB proposed in 2007 optional recourse to the terminology of the Statement of Financial Position in place of what used to be the Balance Sheet; the French translation of "Statement of Financial Position could be" "Situation financière."

It is interesting to put into perspective this information and the accounting elements related to speculation identified by Proudhon that are presented above.

- First of all, the extent of indebtedness within the liabilities should be noted. Shareholders' funds represent only slightly more than 48% of liabilities. The open capital account is thus in full operation. In other words, the equity funds raised are insufficient in relation to the requirements. The shareholders benefit, it should be noted however, from a considerable return, thus making the case for leverage via borrowing. Distributions to shareholders amount to 6,025 + 17,600 in respect of interest and dividends, thereby representing 23,625 from a profit of 31,308.
- Then we should examine the depreciation. It appears weak: 360 in relation to fixed assets of 302,000. Proudhon also cites this amount on page 314 of his work and uses it as an example of misleading accounting practices. However, one must remain prudent. Maintenance expenditure is implemented and accounted for in different accounting sections. The question of fixed capital cannot therefore be analysed easily here since the company is viable, except if one concludes that the depreciation charged here is very much less than that which it should be.

_______________________